\documentclass[11pt]{article}
\usepackage{blois,epsfig}

\def\be{\begin{equation}}
\def\ee{\end{equation}}
\def\bea{\begin{eqnarray}}
\def\eea{\end{eqnarray}}

\def\gappeq{\mathrel{\rlap {\raise.5ex\hbox{$>$}} {\lower.5ex\hbox{$\sim$}}}}
\def\lappeq{\mathrel{\rlap{\raise.5ex\hbox{$<$}} {\lower.5ex\hbox{$\sim$}}}}

\def\npb#1#2#3{    {\it Nucl. Phys. }{\bf B #1} (#2) #3}
\def\plb#1#2#3{    {\it Phys. Lett. }{\bf B #1} (#2) #3}

\def\prep#1#2#3{   {\it Phys. Rep. }{\bf #1} (#2) #3}
\def\prl#1#2#3{    {\it Phys. Rev. Lett. }{\bf #1} (#2) #3}

\def\ppnp#1#2#3{   {\it Prog. Part. Nucl. Phys. }{\bf #1} (#2) #3}

\def\zpc#1#2#3{    {\it Z. Phys. }{\bf C #1} (#2) #3}

\def\ibid#1#2#3{   {\it ibid. }{\bf #1} (#2) #3}

\begin{document}
\vspace*{-1cm}
\phantom{hep-ph/***}

\hfill{RM3-TH/12-10~~~~~~~~~~CERN-PH-TH/2012-157~~~~~~~~~~~}


\vspace*{4cm}
\title{THE SM AND SUSY AFTER THE 2011 LHC RESULTS}

\author{ G. ALTARELLI }

\address{Dipartimento di Fisica `E.~Amaldi', Universit\`a di Roma Tre
and INFN, Sezione di Roma Tre,\\ I-00146 Rome, Italy\\ and \\ CERN, Department of Physics, Theory Unit, 
\\CH--1211 Geneva 23, Switzerland}

\maketitle
\abstracts{
We present a short review of the LHC results at 7 TeV and their implications on the Standard Model (SM) and on its Supersymmetric (SUSY) extension. In particular we discuss the exclusion range for the SM Higgs mass, the tantalizing hint of an excess at $m_H \sim 125$ GeV, the negative results of searches for SUSY particles (as well as for any other new physics) and the present outlook}

\section{The first LHC results}

The main LHC results so far (with more than 5.5 $fb^{-1}$ of integrated luminosity collected by each large experiment at 7 TeV) are listed here, as presented at Moriond 2012.

1) A robust exclusion interval for the SM Higgs has been established which greatly extends the previous range; precisely at 95 $\%$ c.l. the excluded intervals of mass are
ATLAS: 110 -117.5, 118.5 -122.5, 129 - 539 GeV and CMS: 127.5 - 600 GeV  (note that ATLAS	 also	excludes at 95$\%$ c.l. a large part of the mass range 110-122.5 GeV, while CMS has some excess in that region). In addition, there is some tantalizing indication for $m_H \sim  125$ GeV. In this respect, what is encouraging is that an excess is seen in the $\gamma \gamma$ mass distribution both in ATLAS (2.8 $\sigma$ at 126.5 GeV) and CMS (2.9 $\sigma$ at 125 GeV). Also there is a hint for $ZZ \rightarrow 4l^\pm$ in ATLAS (2.1$\sigma$ at ~125 GeV: 3 events) and the Tevatron reports a small excess spread over a large interval in $b \bar b$ and WW (2.7 $\sigma$ in 115 -135 GeV). These accumulations are all compatible with $m_H \sim  125$ GeV. Further encouragement has been missed because in CMS a possible hint in $ZZ \rightarrow 4l^\pm$ is at a different mass (2.1$\sigma$ at ~119.5 GeV: 3 events) and in ATLAS the number of WW events is less than expected. Overall the evidence for  $m_H \sim  125$ GeV could still evaporate and we need to wait for the outcome of the 2012 run, that, with an expected additional integrated luminosity of 15 $fb^{-1}$ per experiment at 8 TeV, should either definitely confirm or exclude $m_H \sim  125$ GeV. 

2) No evidence of new physics has been found although a big chunk of
new territory has been explored.

3) Important results on B and D decays have been obtained mainly by LHCb, whose performance has been exceedingly good (but also on some issues by ATLAS and CMS), e.g. $B_s \rightarrow J\psi \phi$, $B_s \rightarrow \mu \mu$, .... CP violation in D decay.... Most of the results go in the direction of the SM. For CP violation in D decay, it could indeed be a sign of new physics but, in view of the uncertainty in the SM prediction, it is difficult to be sure.

\section{The Higgs Problem}

The experimental verification of the Standard Model (SM) \cite{sprin} cannot be considered complete until the predicted physics of the  Higgs sector \cite{djou1} is not established by experiment. Indeed the Higgs problem is really central in particle physics today \cite{wells}. In fact, the Higgs sector is directly related to most of the major open problems of particle physics, like the flavour problem \cite{isid} and the hierarchy problem \cite{gild}, the latter strongly suggesting the need for new physics near the weak scale, which could also clarify the Dark Matter identity. 

It is clear that the fact that some sort of Higgs mechanism is at work has already been established. The W and Z longitudinal degrees of freedom are borrowed from the Higgs sector and are an evidence for it. In fact the couplings of quarks and leptons to
the weak gauge bosons W$^{\pm}$ and Z are indeed experimentally found to be precisely those
prescribed by the gauge symmetry \cite{ewg,AG}.  To a lesser
accuracy the triple gauge vertices $\gamma$WW and ZWW have also
been found in agreement with the specific predictions of the
$SU(2)\bigotimes U(1)$ gauge theory. This means that it has been
verified that the gauge symmetry is unbroken in the vertices of the
theory: all currents and charges are indeed symmetric. Yet there is obvious
evidence that the symmetry is instead badly broken in the
masses. The W or the Z with longitudinal polarization that are observed are not present in an unbroken gauge theory (massless spin-1 particles, like the photon, are transversely polarized). Not only the W and the Z have large masses, but the large splitting of, for example,  the top-bottom quark doublet shows that even the global weak SU(2) is not at all respected by the fermion spectrum. Symmetric couplings and totally non symmetric spectrum is a clear signal of spontaneous
symmetry breaking and its implementation in a gauge theory is via the Higgs mechanism. The big remaining questions are about the nature and the properties of the Higgs particle(s). The LHC has been designed to solve the Higgs problem.

And indeed the SM Higgs is close to be observed or excluded!
Either the SM Higgs is very light ($ \lappeq$ 128 GeV)
or rather heavy (i.e. $ \gappeq$ 600 GeV). The range $m_H$ = 122 - 128 GeV, where possibly there is a signal, is in agreement
with precision tests, compatible with the SM (the data are in fair agreement with the SM Higgs cross-sections \cite{coupl}) and also with
the SUSY extensions of the SM. Actually,   
$m_H \sim  125$ GeV is what one expects from a direct interpretation
of EW precision tests: no fancy conspiracy with new physics
to fake a light Higgs while the real one is heavy. On the contrary, 
$m_H \gappeq$  600 GeV would point to the conspiracy alternative (but no conspirators have been found nearby!). Thus there is really a great suspense on the LHC run this year. 

What if the evidence $m_H \sim  125$ GeV evaporates in 2012? Can we do without the Higgs? Suppose we take the gauge symmetric part of the
SM and put masses by hand.
What is the fatal problem at the LHC scale?
The most immediate disease that needs a solution is that
in the absence of a Higgs particle or of an alternative mechanism, violations of unitarity appear in scattering amplitudes involving longitudinal gauge bosons (those most directly related to the Higgs sector) at energies in the few TeV range \cite{unit}.
A crucial question for the LHC is to identify the mechanism that avoids the unitarity violation: is it one or more Higgs bosons or some new vector boson (like additional gauge bosons $W^\prime$, $Z^\prime$ or Kaluza-Klein recurrences or resonances from a strong sector)?
Thus something must happen at the few TeV scale! It is not possible that neither the Higgs nor new physics are present at the Electro-Weak (EW) scale (the only caution is whether the LHC can completely explore the EW scale).

It is well known that there are theoretical bounds on the Higgs mass valid if one assumes that the SM, with only one Higgs doublet, is valid up to a large energy scale $\Lambda$ where eventually new physics appears. An upper limit on $m_H$ (with mild dependence
on $m_t$ and $\alpha_s$) is obtained, as described in \cite{hri}, from the requirement that no Landau pole appears, up to the scale $\Lambda$, in the Higgs quartic coupling $\lambda$, or in simpler terms, that the perturbative description of the theory remains valid up to  $\Lambda$. The Higgs mass enters because it fixes the initial value of the quartic Higgs coupling $\lambda$ in the running from the EW scale up to $\Lambda$. Even if $\Lambda$ is as small as ~a few TeV the limit is well within the LHC range $m_H < 600-800~$GeV and becomes $m_H < 180~$GeV for $\Lambda \sim M_{Pl}$. This upper limit on the Higgs mass in the SM has played a crucial role in the LHC design whose mission requires that the whole allowed range is within reach of the machine. A lower limit on
$m_H$ is derived from the requirement of vacuum stability \cite{cab}, i.e. that the quartic Higgs coupling $\lambda$ does not turn negative in its running up to $\Lambda$ (if so the energy would become negative and unbound at large absolute values of the field).  Actually, in milder form, one can tolerate a moderate instability, compatible with the present age of the Universe  \cite{isi}.  A recent thorough reanalysis of this issue \cite{stru1} has concluded that, given the experimental values of $m_t$ and $\alpha_s$, for $\Lambda \sim M_{GUT}-M_{Pl}$ the stability bound is very close to $m_H = 130$ GeV . The value $m_H \sim  125$ GeV would imply that, in the absence of new physics, our Universe becomes metastable at a scale $\Lambda \sim 10^{10}$ GeV. But the lifetime of our vacuum, for scales up to the Planck mass, would be larger than the age of the Universe. The SM remains viable with some early Universe implications. The vacuum could be stabilized by very little
additional new physics (like, for example a heavy singlet S with a large VEV below the metastability scale \cite{stru2}). Large Majorana neutrino masses can also have an impact on the running \cite{stru2}. On the basis of this discussion we can conclude that a 125 GeV Higgs would be nearly perfect for a pure and simple SM up to $M_{Pl}$, just a little bit below the optimal range $130 \lappeq m_H \lappeq 180$ GeV. Incidentally, the possibility that $m_H \sim  130$ GeV, so that the SM becomes unstable precisely at around the Planck mass, and its implications have been studied in the literature \cite{wet}.

\section{Outlook on Avenues beyond the Standard Model}

No signal of new physics has been
found neither in EW precision tests nor in flavour physics. Given the success of the SM why are we not satisfied with that theory? Why not just find the Higgs particle,
for completeness, and declare that particle physics is closed? As well known, the reason is that there are
both conceptual problems and phenomenological indications for physics beyond the SM. On the conceptual side the most
obvious problems are the proliferation of parameters, the puzzles of family replication and of flavour hierarchies, the fact that quantum gravity is not included in the SM and the related hierarchy problem. Among the main
phenomenological hints for new physics we can list the constraints from coupling constant merging in Grand Unified Theories (GUT's), Dark Matter, neutrino masses (explained in terms of L non conservation), 
baryogenesis and the cosmological vacuum energy (a gigantic naturalness problem).
The computable evolution with energy
of the effective gauge couplings clearly points (better in SUSY than in the SM) towards the unification of the electro-weak and strong forces at scales of energy
$M_{GUT}\sim  10^{15}-10^{16}~ GeV$ which are close to the scale of quantum gravity, $M_{Pl}\sim 10^{19}~ GeV$.  One is led to
imagine  a unified theory of all interactions also including gravity (at present superstrings provide the best attempt at such
a theory). Thus GUT's and the realm of quantum gravity set a very distant energy horizon that modern particle theory cannot
ignore. Can the SM without new physics be valid up to such large energies? Indeed, some of the SM problems could be postponed to the more fundamental theory at the Planck mass. For example, the explanation of the three generations of fermions and the understanding of fermion masses and mixing angles can be postponed. But other problems must find their solution in the low energy theory. In particular, the structure of the
SM could not naturally explain the relative smallness of the weak scale of mass, set by the Higgs mechanism at $
1/\sqrt{G_F}\sim  250~ GeV$  with $G_F$ being the Fermi coupling constant. This so-called hierarchy problem \cite{gild}  is due to the instability of the SM with respect to quantum corrections. This is related to
the
presence of fundamental scalar fields in the theory with quadratic mass divergences and no protective extra symmetry at
$\mu=0$, with $\mu$ the scalar mass. For fermion masses, first, the divergences are logarithmic and, second, at
$m=0$ an additional symmetry, i.e. chiral  symmetry, is restored. Here, when talking of divergences, we are not
worried of actual infinities. The theory is renormalizable and finite once the dependence on the cut-off $\Lambda$ is
absorbed in a redefinition of masses and couplings. Rather the hierarchy problem is one of naturalness. We can look at the
cut-off as a parameterization of our ignorance on the new physics that will modify the theory at large energy
scales. Then it is relevant to look at the dependence of physical quantities on the cut-off and to demand that no
unexplained enormously accurate cancellations arise. 

The hierarchy problem can be put in less abstract terms: loop corrections to the Higgs mass squared are
quadratic in the cut-off $\Lambda$. The most pressing problem is from the top loop (the heaviest particle, hence the most coupled to the Higgs).
If we demand that the correction does not exceed the light Higgs mass indicated by the precision tests, $\Lambda$ must be
close, $\Lambda\sim o(1~TeV)$. So a crucial question for the LHC to answer is: what damps the top loop contribution? Similar constraints arise from the quadratic $\Lambda$ dependence of loops with gauge bosons and
scalars, which, however, lead to less pressing bounds. So the hierarchy problem demands new physics to be very close. Actually, this new physics must be rather special, because it must be
very close, yet its effects are not clearly visible in EW precision tests (the "LEP Paradox" \cite{BS}) now also accompanied by a similar "flavour paradox" \cite{isid}. Examples  \cite{CGrojean} of proposed classes of solutions
for the hierarchy problem are SUSY, technicolor, "Little Higgs" models, extra dimensions, effective theories for compositeness etc or the alternative, extreme, point of view given by the anthropic solution. In the following, after a comment on the anthropic route, I will discuss the quest for SUSY in some detail, while the alternative solutions to the hierarchy problem will be considered in the companion presentation by Mariano Quiros \cite{qui}.

\section{An extreme solution: the anthropic way } 

The observed value of the cosmological constant $\Lambda$ poses a tremendous, unsolved naturalness problem \cite{tu}. Yet the value of $\Lambda$ is close to the Weinberg upper bound for galaxy formation \cite{We}. Possibly our Universe is just one of infinitely many bubbles (Multiverse) continuously created from the vacuum by quantum fluctuations. Different physics takes place in different Universes according to the multitude of string theory solutions \cite{doug} ($\sim 10^{500}$). Perhaps we live in a very unlikely Universe but the only one that allows our existence \cite{anto},\cite{giu}. I find applying the anthropic principle to the SM hierarchy problem somewhat excessive. After all one can find plenty of models that easily reduce the fine tuning from $10^{14}$ to $10^2$: why make our Universe so terribly unlikely? If to the SM we add, say, supersymmetry, does the Universe become less fit for our existence? In the Multiverse there should be plenty of less fine tuned Universes where more natural solutions are realized and yet are suitable for our living. By comparison the case of the cosmological constant is a lot different: the context is not as fully specified as the for the SM (quantum gravity, string cosmology, branes in extra dimensions, wormholes through different Universes....). While I remain skeptical I would like here to sketch one possibility on how the SM can be extended in agreement with the anthropic idea.  If we ignore completely the hierarchy problem and only want to reproduce the most compelling data that demand new physics beyond the SM, a possible scenario is the following one. The SM is to be completed by a light Higgs and no other new physics is in the LHC range (how sad!) except perhaps a $Z'$, for example a $Z'_{B-L}$. In particular there is no SUSY in this model. At the GUT scale of $M_{GUT} \gappeq 10^{16}$ GeV the unifying group is $SO(10)$, broken at an intermediate scale, typically $M_{int} \sim  10^{10}-10^{12}$ down to a subgroup like the Pati-Salam group $SU(4)\bigotimes SU(2)_L \bigotimes SU(2)_R$ or some other one \cite{mal}. Note that unification in $SU(5)$ would not work because we need a group of rank larger than 4 in order to allow for a two step (at least) breaking needed, in the absence of SUSY,  to restore coupling unification and to avoid a too fast proton decay. The Dark Matter problem should be solved by axions \cite{axi}. Lepton number violation, Majorana neutrinos and the see-saw mechanism give rise to neutrino mass and mixing. Baryogenesis occurs through leptogenesis \cite{buch}. One should one day observe proton decay and neutrino-less beta decay. None of the alleged indications for new physics at colliders should survive (in particular even the claimed muon (g-2) \cite{amu} discrepancy should be attributed, if not to an experimental problem, to an underestimate of the theoretical errors or, otherwise, to some specific addition to the above model \cite{stru3}). This model is in line with the non observation of $\mu \rightarrow  e \gamma$ at MEG \cite{meg}, of the electric dipole moment of the neutron \cite{nedm}
etc. It is a very important challenge to experiment to falsify this scenario by establishing a firm evidence of new physics at the LHC or at another "low energy" experiment.

\section{Supersymmetry}

In the limit of exact boson-fermion symmetry \cite{Martin} the quadratic divergences of bosons cancel so that
only log divergences remain. However, exact SUSY is clearly unrealistic. For approximate SUSY (with soft breaking terms),
which is the basis for all practical models, $\Lambda$ is essentially replaced by the splitting of SUSY multiplets. In particular, the top loop is quenched by partial cancellation with s-top exchange, so, to limit the fine-tuning the s-top cannot be too heavy. The existing limits on SUSY particles (even before the LHC), EW precision tests,
success of the Cabibbo-Kobayashi-Maskawa theory of quark mixing and of CP violation, absence of Flavour Changing Neutral Currents, all together,
impose sizable fine tuning particularly on
minimal realizations.
Yet SUSY is a completely specified, consistent, computable
model, perturbative up to $M_{Pl}$. Important phenomenological indications in favour of SUSY are that coupling unification takes place with greater accuracy in SUSY than in the SM and that proton decay bounds are not in contradiction with the predictions. Grand Unification (GUT's) and SUSY go very well together: this is unique among new physics models. Other non standard models \cite{CGrojean,qui} (little Higgs, composite Higgs, Higgsless....)
all become strongly interacting and non perturbative
at a multi-TeV scale. Two Higgs doublets are expected in SUSY \cite{djou2}. The EW symmetry breaking can be triggered by the $H_u$ mass becoming negative at low energy in the running down from the GUT scale, due to the large top Yukawa coupling. SUSY offers a good Dark Matter candidate: the neutralino
(actually more than one candidate, e.g. also the gravitino).
In summary SUSY remains the reference model for new physics. But the negative result of the search for SUSY at the LHC, where a big chunk of new territory has been explored in the last year run, has imposed new strong constraints on SUSY models. And the hint of $m_H = 125$ GeV, if confirmed, does even more restrict the parameter space of these models ($m_H = 125$ GeV is a bit too heavy: near the upper bound on $m_H$ in the MSSM).

Even the Minimal SUSY Model (MSSM) \cite{Martin} has more than 100 parameters (mostly from the SUSY soft breaking terms).
Simplified versions with a drastic reduction of parameters
are used for practical reasons, e.g. the
CMSSM, where C stands for Constrained, or mSUGRA, i.e. minimal SuperGravity (often the two names are confused): with universal gaugino and scalar soft terms
at the GUT scale, the set of parameters is drastically reduced down to $m_{1/2}$, $m_0$, $A_0$ (the s-top mixing parameter), $\tan{\beta}$ and sign($\mu$). Similarly in the Non Universal Higgs Mass models NUHM1,2: masses for Hu, Hd (1 or 2 masses) different from $m_0$
are added. It is only these oversimplified models that are now cornered. A more flexible setup  but, apparently still manageable, is the MSSM
with CP and R conservation (pMSSM: p for phenomenological) \cite{pmssm} in terms of 19 parameters ($M_A$, $\tan{\beta}$, 3 gaugino masses,  3 mixing parameters $A_u$, $A_d$, $A_e$, $\mu$ and 10 s-fermion masses, with degenerate first 2 generations) recently studied in several works. 

Many different new physics signatures have been searched at the LHC at 7 TeV with no positive outcome in a variety of channels involving combinations of charged leptons, jets and missing energy. All kinds of models for new physics can be compared with these data, not only SUSY. For SUSY the resulting limits depend on the assumptions on the spectrum, but, in the CMSSM, generically imply that gluinos and degenerate s-quarks are heavier than 500 - 1000 GeV. In addition to these limits the impact of $m_H \sim$ 125 GeV on SUSY models is important \cite{mam}. For example, minimal models with gauge mediation or anomaly mediation are disfavoured \cite{arb}
(predict $m_H$ too light) although some versions, like gauge mediation with extra vector like matter \cite{endo},
could still work. Specific models that give up naturalness but remain predictive like split SUSY or heavy SUSY have seen their allowed domain restricted \cite{giu}. Gravity mediation \cite{grmed} is in better shape but CMSSM, mSUGRA, NUHM1,2
are only marginally consistent and need s-quarks heavy, $A_t$ large and lead to tension with the muon (g-2). In fact the muon magnetic moment would point to light SUSY, more precisely to light EW gauginos and s-leptons. This type of light SUSY would also improve the EW precision fit (by predicting a heavier $m_W$ than the SM for the experimental value of $m_t$ and a light Higgs).  Several groups (for example, see \cite{OBuch}) have repeated the fit to EW precision tests in the CMSSM, also including the additional data on the muon $(g-2)$, the Dark Matter relic density and rare $b\rightarrow s \gamma$ decay modes. Before the LHC results the promising outcome of this exercise was that the central value of the lightest Higgs mass $m_H$ went up (in better harmony with the bound from direct searches) with moderately large $\tan{\beta}$ and relatively light SUSY spectrum \cite{OBuch}. After the LHC bounds one finds that the best fit Higgs mass is 125 GeV only if the result on the muon (g-2) is removed from the fit, while, with the (g-2) included, the best fit Higgs mass value is 119 GeV. In other words, in the CMSSM there is a sizable tension between the muon (g-2) and $m_H \sim$ 125 GeV. 
Also normally too much Dark Matter is predicted in the CMSSM or mSUGRA for $m_H =$ 125 GeV. In comparison, the upper limit on $m_H$ is larger in the pMSSM: $m_H \lappeq$ 135 GeV \cite{pmssm}.

The problem with SUSY is that one expected its discovery already at LEP2 on the basis of complete naturalness applied to minimal models. With the recent LHC data ever increasing fine tuning appears to be needed in the minimal versions. However less fine tuning is necessary if non minimal models are assumed.  One must go beyond the CMSSM, mSUGRA, NUHM1,2. And indeed there is still plenty of room for more sophisticated versions of
SUSY as a solution to the hierarchy problem. The simplest new ingredients that are studied at present are either
heavy first 2 generations \cite{heavy1,heavy2} and/or an extra Higgs singlet \cite{exs}. 

The first option is still within the MSSM framework. Note that, on the one hand, it is mostly gluinos and 1-2 generation s-quarks that are affected by the LHC limits
while EW s-particles and s-tops are less constrained. On the other hand, what is really needed for naturalness in the MSSM \cite{heavy2} is that the s-tops (they directly enter at one loop in the radiative corrections to the Higgs mass), their isospin partners the s-bottoms, as well as the lightest higgsino (related to the $\mu$ parameter), and also gluinos (that contribute, with a strong coupling, in the radiative corrections at two loops) must be relatively light (below, say, 1 TeV). As remarked already long ago \cite{heavy1} an inverted s-quark spectrum with heavier 1st-2nd and lighter 3rd generation s-quarks has several advantages in flavour and CP violation problems. This option has been widely reanalysed recently in the present context. If gluinos are forced to only decay into final states involving s-tops or s-bottoms, their mass limits are considerably less stringent. Similarly the present lower limit on the lightest s-top mass is a few hundred GeV.

By adding an extra singlet Higgs \cite{exs} one goes beyond the MSSM. 
In a promising class of models a singlet Higgs S is added with coupling $\lambda S H_uH_d$.
The $\mu$ term arises from the S Vacuum Expectation Value (VEV) and the $\mu$ problem is solved in that the S VEV can naturally be of order of the soft terms that break SUSY. Mixing with S can modify the Higgs mass and couplings at tree level. In particular, the restrictions on the Higgs mass, valid at tree level in the MSSM that demand substantial corrections from loops, can be relaxed (no need of large s-top mixing, less fine tuning). The new coupling $\lambda$ grows with the scale. If we impose that the theory remains perturbative up to $M_{GUT}$ then we must have 
$\lambda \lappeq \sim 0.7$. This is the case of the NMSSM (Next to Minimal SSM). For $m_H \sim 125$ GeV larger values of $\lambda$ allow for lighter s-tops, no large s-top mixing and much less fine tuning. For $\lambda \sim 1 - 2$ we are in the so-called $\lambda$-SUSY regime (for $\lambda \gappeq 2$ the theory becomes non perturbative already at ~10 TeV). The fine tuning can be really reduced to a few percent even with a s-top of mass above 1 TeV. The presence of an extra Higgs singlet adds one more neutral scalar particle to the spectrum. After symmetry breaking the mixing between S and the doublet Higgs leads to two eigenstates of mass that replace the single lightest Higgs h (for not too large $\lambda$ a 2 by 2 matrix mixing approximation is valid while for $\lambda \gappeq 0.7$ the full mixing matrix must be considered). The state at 125 GeV could be the lightest, but it is not excluded that at 125 GeV the heaviest of the two
is seen while the lightest escaped detection at LEP \cite{ell}. In fact the mixing also modifies the couplings and may be that the lightest eigenstate has suppressed couplings to gauge bosons. In this case the heavier one at 125 GeV would get enhanced couplings to gauge bosons. Indeed there is a tenuous indication that the 125 GeV state may have a slightly enhanced coupling to $\gamma \gamma$.

\section{Conclusion}

The most exciting result of the 2011 LHC run is that the SM Higgs is close to be observed or excluded!
The present, very solid, exclusion ranges for the SM Higgs have much restricted the mass interval for the SM Higgs: either the SM Higgs is very light (115 - 128 GeV)
or very heavy (i.e. $ \gappeq$ 600 GeV). 
The range $m_H$ = 122 - 128 GeV where some excess is observed is in agreement
with precision tests, compatible with the SM and also with
the SUSY extensions of the SM. This hint is very exciting but could still disappear with more statistics. Thus the outcome of the 2012 LHC run at 8 TeV is of extreme interest for particle physics.

The search for new physics is the other big issue. No signals have shown up so far in spite of the many channels explored and of the large slice of parameter space that has been for the first time explored. Optimistic expectations of an early success have been deceived. But the LHC experiments are just at the start and
larger masses can be reached in 2012
and even more in the 14 TeV phase.  Still supersymmetry remains the standard way beyond the SM.  It is true that we could have expected the first signals of SUSY already at
LEP, based on naturality arguments applied to the most minimal models like the CMSSM or mSUGRA. But the general MSSM is still very much viable, for example in the versions with heavy 1 - 2 generation s-quarks \cite{heavy2}. Among non minimal models the most studied possibility are based on the addition of an extra singlet S to the Higgs sector \cite{exs} (NMSSM and $\lambda$ - SUSY). The absence of SUSY signals has also stimulated the development of new ideas like those of extra dimensions and composite Higgs (discussed in the talk by M. Quiros \cite{qui}). The extreme anthropic proposal that naturalness could be irrelevant for the very particular physics that is valid in our exceptional Universe, just one among many in the Multiverse, is boosted now by the absence of new physics signals at the LHC. Only experiment can choose among these and other possibilities. Supersymmetry? Compositeness? Extra dimensions? Anthropic? We shall see!

\vspace{0.5cm} 
I am very grateful to J. Tran Thanh Van and the Organisers of the 2012 Rencontres de Moriond, in particular B. Klima and B. Pietrzyk, for inviting me to give this talk. My related research has been funded by the COFIN program (PRIN 2008), the INFN- Roma Tre, and by the European Commission, under the networks ``LHCPHENONET'' and ``Invisibles''.

\end{document}